\documentclass[twocolumn,prl,floatfix,showpacs]{revtex4}

\usepackage{bm}
\usepackage{amsmath,amssymb}
 \usepackage{palatino}
 \usepackage[osf]{mathpazo}
 \usepackage[dvips]{graphicx}
 \usepackage[dvips]{color}
 \usepackage{pslatex}

\def\ber{\begin{eqnarray}}
\def\eer{\end{eqnarray}}
\def\be{\begin{equation}}
\def\ee{\end{equation}}
\def\bea{\begin{eqnarray}}
\def\eea{\end{eqnarray}}

\begin{document}

\title{Disclination-mediated thermo-optical response in nematic glass sheets}
\author{Carl D. Modes$^1$, Kaushik Bhattacharya$^2$, and Mark Warner$^1$} \affiliation{$^{1}$ Cavendish Laboratory, University of Cambridge, Madingley Road, Cambridge CB3 0HE, UK \\  $^{2}$ Division of Engineering and Applied Science, California Institute of Technology, Pasadena, CA, 91125, USA}
\date{\today}

\begin{abstract}
Nematic solids respond strongly to changes in ambient heat or light,
significantly differently parallel and perpendicular to the
director. This phenomenon is well characterized for uniform director
fields, but not for defect textures.  We analyze the elastic ground
states of a nematic glass in the membrane approximation as  a
function of temperature for some disclination defects with an eye
towards reversibly inducing three-dimensional shapes from flat
sheets of material, at the nano-scale all the way to macroscopic
objects, including non-developable surfaces.  The latter offers a
new paradigm to actuation via switchable stretch in thin systems.
\end{abstract}
 \maketitle

Nematic glasses (densely crosslinked networks) \cite{Mol:91} are
solids with a natural state of elongation along their director, and
contraction perpendicular, that depends on their orientational
order.  Accordingly they suffer large, reversible length change with
heating \cite{Mol:05}, illumination \cite{Harris:05,vanoosten:07},
solvent uptake, pH change and any other stimulus that causes order
change. For glasses these strains can be 2-3\% and of opposite sign
(without conserving volume) along and perpendicular to the director.
They can then exhibit spectacular effects such as large bend
resulting from gradients of stimuli \cite{Ikeda:03} (light, solvent)
or from uniform stimuli (like temperature or weakly absorbing light)
but with a director gradient through, for instance, the section of a
sheet or cantilever \cite{Mol:05,Harris:05,vanoosten:07} (usually
twist or splay-bend). Glasses are distinguished from elastomers by
being so heavily cross-linked that the director field only changes
as a result of convection due to material shape change from the
elastic strain.  Nematic glass directors do not rotate relative to
the matrix as in nematic elastomers -- they are \cite{Harris:05}
conventional, uniaxial elastic solids, with moduli about $\times
10^4$ higher than those of elastomers.  We are most concerned with
the thermally or optically inspired length changes which we take to
be by a factor $\lambda$ along the director and by $\lambda^{-\nu}$
perpendicular to the director (where $\nu$ fulfills the role of a
thermal or optical Poisson ratio, were the deformations to be
infinitesimal).  Elastic energy cost would be associated with
imposed changes away from these new natural shapes -- we find
deformation fields matching these thermo/optical changes and thus of
zero energy in the  membrane limit we take.

Director fields can be established in the nematic liquid progenitor
phase before crosslinking and are permanently recorded in the solid
state achieved after linkage.  In fact complex, 3-D director fields
for subtle mechanical response can be achieved in nematic glasses
\cite{Serrano:08} via holography and surface preparation.  From the
point of view of device design, the ability to "write" an initial
director field into a solid so that it distorts in a predictable
(and reversible way) into a new shape with applied temperature or
light would be ground-breaking.  In particular, the ability to take
flat, thin sheets of material and turn them into prescribed,
potentially complicated, non-developable shapes -- at nearly any
length scale -- is highly sought after.  Accordingly, we consider
here the elastic response of thin sheets of nematic glass in the
 membrane approximation.
We are interested in thin sheets where the director is uniform
through the thickness.  In this situation, a change in nematic order
gives rise to in-plane stretches (or contractions).  If the initial
director field is not homogeneous, then this change of nematic order
gives rise to inhomogeneous changes of stretch that may or may not
be compatible.  A key observation here is that thin sheets can
accommodate many more potential inhomogeneous changes of stretch
compared to bulk 3D specimens by possibly deforming out of plane.
Further, in thin sheets, the energy and forces associated with
membrane mode (in-plane stretch) scales as the thickness while those
associated with bending mode (differential in-plane stretch) scales
as the third power of thickness.   So the bending energy is
negligible compared to the in-plane stretches.  All of this leads to
a new paradigm for actuation \cite{BJ:05}:  If we are able to find a
director arrangement that leads to stretches that are incompatible
in bulk but compatible in sheets, then the change of order can be
used to generate out-of-plane deformations with a very large
blocking force (one that scales with thickness). This is in contrast
with small blocking forces (third power of thickness) associated
with designs involving bending cantilevers. In this Letter, we show
that +1 disclinations are indeed such arrangements.
\begin{figure}[!b]
\centerline{\includegraphics[width=8cm]{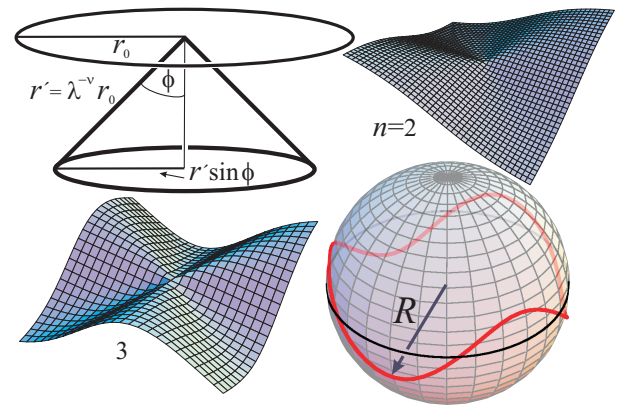}}
\caption{A flat nematic glass sheet with an azimuthal +1
disclination heats to a cone, or cools to ``anticones" ($n$).}
\label{fig:anticones}
\end{figure}

Disclination defect textures are not extensively studied,
experimentally or theoretically, in nematic solids.  However it is
known that the energy minimizing  deformation associated locally
with changes in their nematic order is not compatible in some 3-D
disclinations \cite{Fried}. We analyze the mechanical response  and
elastic ground state of the most experimentally accessible
disclination defect in 2-D, that with topological charge +1.  We
argue that it is a powerful way to induce shape change, specifically
through the introduction of a point of localised Gaussian curvature.
Fig.~\ref{fig:anticones} shows examples of initially flat, defected
sheets after heating or cooling.

We illustrate some ways director fields with this disclination
charge could be used in concert with system geometry for
applications, such as sharp bending, twisting or forming
non-developable (anticlastic) surfaces and cantilevers.

Shape response to nematic defects has been investigated
theoretically in fluid systems \cite{Uchida:02,Frank:08} where the
dominant influences include surface tension, Frank elasticity and
order-curvature coupling. Our solids are in the opposite limit --
elastic stresses dominate over Frank elasticity for length scales
greater than a nematic penetration depth for solids $\xi =
\sqrt{K/\mu}$, where $\mu$ and $K$ are  the shear modulus and a
Frank constant. ($\xi \sim 10^{-10}$m for a glass.)  Also, in our
problem surface energies  play no role, and order resists
mechanically induced change.

Response closely analogous to ours has been analyzed by Ben Amar
{\it et al} in the elasticity of  botanical systems when anisotropic
growth creates internal stresses and forces planar systems into the
third dimension \cite{Ben-Amar:08a,Ben-Amar:08b} as cones or
``e-cones" (our anti-cones). They and we deal with localized
Gaussian curvature and hence localized elastic stretch. Such stress
intensification occurs in folding and crumpling, see Witten's review
\cite{Witten:07}. However, our cones and anti-cones are simple, not
``d-cones" \cite{Witten:07}, and our systems naturally take up such
shapes rather than concentrate stress in response to spatial
crowding. Indeed, except for very weak spontaneous distortions, our
tip extent is of the order of the thickness as in classical simple
cones \cite{Witten:07} -- we return elsewhere to the core/far-field,
bend/stretch energy balance.  Thus our system differs from the cone
sources that generate crumpling.

\begin{figure}[!ht]
\centerline{\includegraphics[width=8cm]{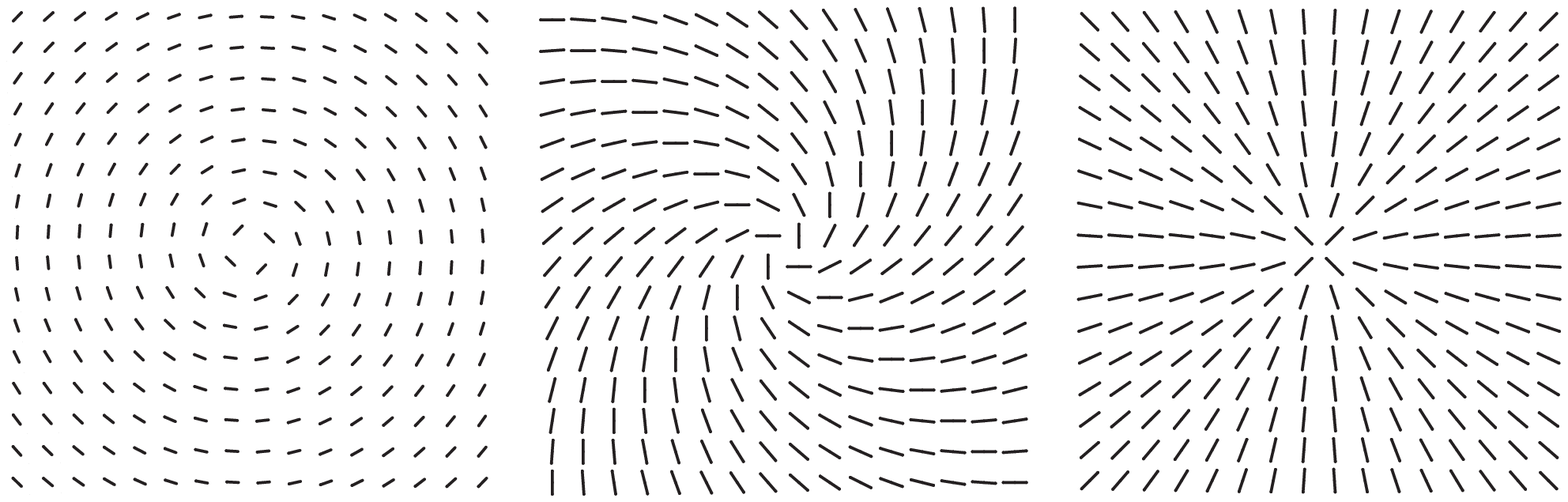}}
\caption{ +1 disclinations -- azimuthal, spiral and radial textures.}
\label{fig:PlusOne}
\end{figure}

We consider +1 disclinations.  In the membrane limit in which we
choose to work, escape in to the third dimension \cite{Meyer_3D:73}
of the director field is not possible except perhaps near the core,
and these disclinations are true topological defects.  For +1
defects, many different textures, though topologically equivalent
(see Fig~\ref{fig:PlusOne}),  differ non-trivially  in their
mechanical response. We  first analyze azimuthal and radial
textures.

Consider a thin sheet of nematic glass whose director field is
azimuthal around a +1 disclination defect, as in the left of Figure
\ref{fig:PlusOne}, and which is flat at some reference temperature,
$T_0$.  As the sample is heated above $T_0$,  the decline in nematic
order will cause a contraction of the natural length along the
nematic directors with a local elongation of the natural length due
to Poisson effects normal to them.  In a free, uniform glass these
natural length changes would be manifested by actual mechanical
strains so that the elastic ground state is achieved.  Since the
chosen director field is circularly symmetric, with integral curves
simply concentric circles centered on the defect, clearly the sheet
has a problem accommodating this change in the natural lengths as
circumferences  shrink while the corresponding natural radii grow.

Fortunately, in the membrane approximation where we may neglect
bending energies, there is an obvious geometric solution that allows
the nematic glass to respond to the imposed thermal strain without
paying the high energetic cost associated with elastic compressions
and expansions relative to this changed state -- a cone.  This may
be seen intuitively by keeping track of the deformation of a circle
of material centered on the defect.  At $T_0$ the sheet of nematic
glass is flat, and the circle maintains the familiar
perimeter-to-radius ratio of $2 \pi$.  However, as the temperature
rises, the perimeter wants to change by a factor of the thermal
deformation gradient along the director, $P \rightarrow P' = \lambda
P$, where here $\lambda < 1$.  Meanwhile, see
Fig.~\ref{fig:anticones},  the material (in-plane) radius is
changing as well due to the thermal/optical Poisson effects
associated with the perimeter's change, $r_0 \rightarrow r' =
\lambda^{-\nu} r_0$. For glasses $\nu$ is in the range 1/3 to 2
\cite{Harris:05}. Together, these transformations imply that upon
heating, a circle of material on the sheet remains circular, but
adopts a new in-material perimeter-to-radius ratio, that of a circle
 enclosing a cone's tip (Fig.~\ref{fig:anticones}, 1st panel):
 \bea
P' = 2 \pi \lambda^{1+\nu} r' &\rightarrow& 2 \pi r' \sin \phi \nonumber\\
\rightarrow \phi(T-T_0) &=& \sin^{-1} \left( \lambda^{1+\nu} \right)
\label{eq:cone-angle}\eea with  $\phi$  the cone opening angle.  One
can think of $r'\sin\phi$ as the embedded radius.  The localized
Gaussian curvature associated with the cone tip is thus
$2\pi(1-\sin\phi)$; circles enclosing the tip of a non-developable
cone no longer have the ratio $2\pi$ of the perimeter to the
in-plane radius, unlike circles on the cone but not enclosing the
tip. Away from the singularity, there is no Gaussian curvature and
hence no shape-induced elastic compressions or extensions. Therefore
our sheet of nematic glass responds to temperatures above $T_0$ by
deforming out of plane, breaking up-down symmetry in the process.
The opening angle varies sensitively with small strains; a 2\%-3\%
contraction with $\nu \sim 2$ achieved over $90^{\rm o}$C
\cite{Mol:05} gives $\phi \sim 70^{\rm o}$--$66^{\rm o}$, such
strains also achievable for modest illuminations in photo-glasses
\cite{vanoosten:07}.

But what if we cool the azimuthal sample below $T_0$, increasing the
nematic order relative to the reference state?  The arguments
relating the local changes in length along and normal to the nematic
director go through unchanged, however now $\lambda > 1$,
invalidating our previous ansatz cone solution -- now the perimeter
is "too long" for the in-plane radius and a more complex deformation
must result.

\begin{figure}[!t]
\centerline{\includegraphics[width=7cm]{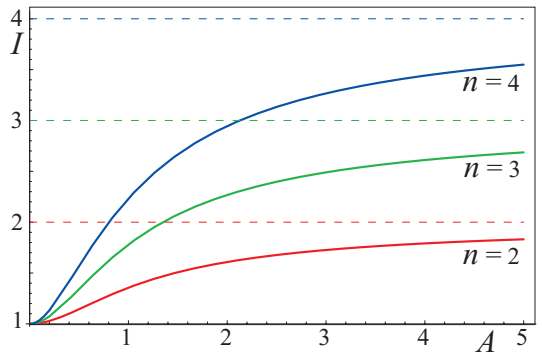}}
\caption{The behavior of $I(n,A)$ for different values of $n$ and as
a function of $A$.  When $A$ is large, $I(n,A)$ approaches $n$.}
\label{fig:limits}
\end{figure}

We consider deformed surfaces where the height varies linearly with
the distance from the center or the defect.  Consider a circle of
radius $r_0$ centered at the defect in the flat undeformed reference
plane.   After deformation, this circle becomes a curve described by
$\{ r(\phi), \phi, h(\phi)\}$ in cylindrical coordinates $(r,\phi,
h)$ in the deformed configuration and parametrized by the azimuthal
angle $\phi$.  If the material is in the minimal energy state
unstretched from its new natural configuration, this curve has to
satisfy two constraints.  First, it has to have a constant distance
$R = \lambda^{-v} r_0$ from the origin (defect) and second, its
length has to be equal to $P = 2 \pi \lambda r_0 = 2 \pi
\lambda^{1+\nu} R$.   Thus, the curve has to lie on a sphere of
radius $R$; see the trajectory in the last panel of Fig.~\ref{fig:anticones}.  Further, since the length of the curve $P$ is greater
than the length $2\pi R$ of the great circle of the sphere, this
curve has to oscillate.   In other words, the deformed surfaces
oscillate azimuthally and these oscillations grow linearly radially
as shown in Fig.~\ref{fig:anticones}.   We call them anti-cones.  We
also note,
\begin{eqnarray}
R &=& r \sqrt{ 1 + \left(\frac{h}{r}\right)^2}\label{radius} \\
P &=& \int_0^{2\pi} d\phi \sqrt{ \left(\frac{dr}{d\phi} \right)^2 +
\left( \frac{dh}{d\phi} \right)^2 + r^2 }\label{perimeter}
\end{eqnarray}
The simplest possibilities for $h$ are
\begin{equation}
h (r; A,n) = Ar \sin n \phi
\end{equation}
for antinodal line angle $\alpha = \tan^{-1}A$ (amplitude in effect;
see Fig.~\ref{fig:anticones}) and integer $n$ (so  that the curve is
closed). Plugging this ansatz into eqn~(\ref{radius}) gives the
relationship between the $r$ coordinate and $\phi$ at constant $R$
needed for the perimeter:
\begin{equation}
r = \left( 1 + A^2 \sin^2 n \phi \right)^{-\frac{1}{2}} R.
\end{equation}
Returning this relation and the form of $h$ to
eqn.~(\ref{perimeter}) for the perimeter gives,  grouping  factors
of $R$ and simplifying, \be P(R) = 2 \pi R I(n,A)\ee where $I(n,A)$
depends only on the scale $A$ and the state  $n$:
 \be I(n,A) = \int_0^1
du \sqrt{\frac{n^2 A^2 \cos^2 2 \pi n u }{(1+A^2 \sin^2 2 \pi n
u)^2} + \frac{1}{1+A^2 \sin^2 2 \pi n u}}\;.\nonumber\ee Connecting
the radius and perimeter  as in eqn~(\ref{eq:cone-angle}) gives here
$I(n,A) = \lambda^{1+\nu}(T-T_0)$; as temperature and hence
spontaneous distortion changes, so does the character (that is, $A$
and $n$) of the anticone. The negative  Gaussian curvature localized
at the apex of the anticone is $2\pi(1-I)$. As appropriate, $I = 1$
for $n=0$ -- the surface is a flat plane.  Otherwise  $I$ ranges
from 1 to $|n|$ for  $A=0\rightarrow \infty$, see
Fig.~\ref{fig:limits}.

The analogous cosine solutions simply give rise to rotated versions
of the same surfaces for all $n\ne 0$,  recovering the conical
solution discussed earlier for $n=0$.  In this case $I(A;n=0)$
ranges from 1 at $A=0$ to 0 at $A=\infty$, as required.  The
behavior of $I$ for  $n\ne 0$ is encouraging -- our trial solutions
yield precisely the geometries that accommodate at zero stretch
energy a cooling of our azimuthal +1 defect below $T_0$. However,
since each individual surface is limited to a maximum
perimeter-to-radius ratio of $2 \pi |n|$, we would expect
interesting transition behavior as cooling leads to strains
requiring ever more crumpled geometries, with transition states
characterized by simple Fourier combinations of the ``pure" surfaces
or, in extreme cases, exotic surfaces that are multiply re-entrant
in $\phi$. Figure~\ref{fig:anticones} shows an $n=3$ anti-cone where
the crumples take up more perimeter than the more slowly varying
$n=2$ anti-cone.  See \cite{Ben-Amar:08b} for an analysis of
large amplitude anti-cones where the functions $h(\phi), r(\phi)$
may not be single valued.  We conjecture that all minimal membrane
energy solutions of these defects are anti-cones.

  A new paradigm for actuation now arises.
Although we are dealing with thin sheets (and later cantilevers) in
the membrane approximation where shape change has low bending
energy, large forces can be exerted by switching on and off Gaussian
curvature via stretch modes that arise if the natural and imposed
geometries are in conflict.

For radial textures, Fig.~\ref{fig:PlusOne}, the roles of the direct
thermal strain and Poisson strain are swapped.  Heating above $T_0$
now requires a shortening of the radial length due to decreased
nematic order while the azimuthal direction expands from the
corresponding Poisson effect -- equivalent to \textit{lowering} the
temperature in the azimuthal texture.  Hence, each of these textures
behaves as the other under the mapping $T-T_0 \rightarrow T_0 - T$.

\begin{figure}[]
\centerline{\includegraphics[width=4cm]{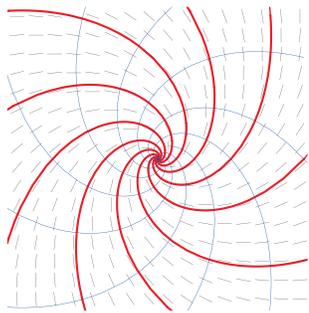}} \caption{A
+1 disclination texture with intermediate $\delta = 45^{\rm o}$. The
integral curves of the director field (heavy, red) and its
complement (light, blue) are logarithmic spirals.}
\label{fig:Spirals}
\end{figure}

Generically a +1 disclination defect has an intermediate angle
$\delta$ of the director with respect to the radial vector from the
defect core.  As a result, the effect of the direct and Poisson
strains are now mixed along circles and radii centered on the
defect.  Curves along which the material feels the maximal effect of
the direct strain and none of the Poisson strain, and vice versa,
are the integral curves of the director field and its normal
complement, respectively.  For +1 textures with $0 < \delta <
\pi/2$, the integral curves are logarithmic spirals instead of
simple circles and radii, Fig.~\ref{fig:Spirals}.
An analysis we present elsewhere shows that the spiral angle
$\delta$ evolves with spontaneous deformation $\lambda$.  Defining
$b = \cot\delta$, one finds that $b \rightarrow b/\lambda^{1+\nu}$.
For instance for heating where $\lambda <1$, then the angle $\delta$
decreases - the spiral tends more to the radial direction.  Also a
mismatch between radius and perimeter as considered above arises:
cones or anticones form depending on whether the radius or perimeter
grows relative to the other.  For instance for an initial $\delta >
\pi/4$ and hence $b$ initially $<1$, the spiral gives a conical
response for $\lambda \in (b^{2/(1+\nu)},1)$ and anticones
otherwise.  The cone angle is $\phi = \sin^{-1}\left[
\frac{\lambda^{2(1+\nu)} + b^2}{1 +
b^2}\frac{1}{\lambda^{1+\nu}}\right]$.
 \begin{figure}[]
\centerline{\includegraphics[width=8.5cm]{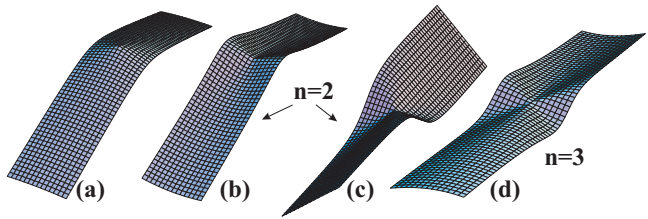}}
\caption{Disclinations in cantilevers also cause unusual thermal or
optical response. (a) a strip cut from a +1 texture conically
deforming (with $\phi = 45^{\rm o}$) bends  with a conical cusp.
 Anti-conical deformations are even richer (here $\alpha =
45^{\rm o}$), with (b) anti-clastic bending ($n=2$, anti-nodes
aligned with the cantilever axes) (c) pure twist (nodes aligned) or
(d) curvature reversal ($n=3$).} \label{fig:Strips}
\end{figure}

There is a new effect, however -- because of the spiral angle of the
director relative to the radial direction, material undergoes
 motion with an azimuthal component.  Furthermore, if $\delta$
and $\lambda$ are such that an out-of-plane deformation into a cone
is required,  then this rotation, combined with the material's
spontaneous choice to form an ``upward"  or ``downward" pointing
cone as the material moves from two to three dimensions, leads to a
spontaneously broken chiral symmetry.  This symmetry breaking does
not occur for the anti-cone solutions, as they do not break up-down
symmetry.

Consider advantageous ways to use such sheets other than as
cone/anti-cone machines. Simplest would be to break the circular
symmetry by cutting out a cantilever of material. Strips cut from
these defected textures can display a wide range of rich behavior,
Fig.~\ref{fig:Strips}: cusped with sharp bending, pure twist,
curvature reversal, and combinations of these depending on the
orientation of the cantilever with the defect. Blocking forces would
depend on the dimensions of the (anti) conical region compared with
the length of the arms. Boundary conditions are very important, and the cut out material must encompass the defect core.
Such active response in strips could be used as a light activated
stirrer, actuator, swimmer, or perhaps as thermally-sensitive, simple machines.

``Frozen-in" disclinated director fields in responsive nematic
glasses are  rich and promising systems.  To exploit them to their
fullest however, will rely on understanding the mechanical response
of all the defect charges, along with how interacting multiple
defects influence the resultant strain-mediated shape change.  Such
 understanding  would allow blueprinting an
arbitrary three-dimensional shape in a flat sheet and switching it
on at will.

CDM, KB and MW acknowledge support from the EPSRC.

\end{document}